\documentclass[reqno,11pt]{amsart}

\usepackage[english]{babel}
\usepackage[utf8]{inputenc}
\usepackage[T1]{fontenc}

\usepackage{amsthm}
\usepackage{physics}
\usepackage{enumitem}
\usepackage{cite}
\usepackage{pdflscape}

\usepackage[widespace]{fourier}

\usepackage{etoolbox} % for '\AtBeginEnvironment' macro
\AtBeginEnvironment{pmatrix}{\everymath{\displaystyle}}
\AtBeginEnvironment{bmatrix}{\everymath{\displaystyle}}

\numberwithin{equation}{section}

\newcommand{\R}{\mathbb{R}}

\newcommand{\Cp}{\mathbb{C}}

\newcommand{\su}{\mathfrak{su}}
\newcommand{\so}{\mathfrak{so}}

\renewcommand{\vec}[1]{\mathbf{#1}}
\renewcommand{\epsilon}{\varepsilon}
\renewcommand{\imath}{\mathrm{i}}
\newcommand{\I}{\mathcal{I}}

\theoremstyle{plain}

\theoremstyle{definition}

\theoremstyle{remark}
\newtheorem{remark}{Remark}[section]
\theoremstyle{remark}

\title{Lax pairs for the discrete reduced Nahm systems}
\author{G. Gubbiotti}
\address{School  of  Mathematics  and  Statistics  F07,  The  University  of  Sydney,  NSW  2006, Australia}
\email{giorgio.gubbiotti@sydney.edu.au}

\date{\today}

\begin{document}

\maketitle

\begin{abstract}
    We discretise the Lax pair for the reduced Nahm systems
    and prove its equivalence with the Kahan--Hirota--Kimura discretisation 
    procedure.
    We show that these Lax pairs guarantee the integrability of the 
    discrete reduced Nahm systems providing an invariant.
    Also, we show with an example that Nahm systems cannot solve
    the general problem of characterisation of the integrability for
    Kahan--Hirota--Kimura discretisations.
\end{abstract}

\section{Introduction}

W. Nahm in 1982 \cite{Nahm1982} introduced a model for self-dual
multimonopoles in terms of three coupled matrix differential equations:
\begin{equation}
    \dot{T}_{i} = [T_{j},T_{k}],
    \quad T_{i}=T_{i}\left( t \right)\in M_{N,N}\left( \Cp \right),
    \label{eq:nahm}
\end{equation}
where the indices $i,j,k$ are cyclic permutations of the set
$\left\{ 1,2,3 \right\}$, and $N$ is a positive integer.
The system of three equations \eqref{eq:nahm} nowadays called Nahm's equations.
%Consider three $N\times N$ matrices  satisfying
%the following differential equations:

In \cite{Hitchinetal1995} some special cases of Nahm's equations
with particular symmetries were studied in connection with
the theory of monopoles.
The obtained systems of coupled two-dimensional differential
equations are known as the \emph{reduced Nahm systems}:
\begin{subequations}
    \begin{gather}
        \dot{x}= 2x^2+\frac{y^2}{8}, 
        \quad 
        \dot{y} = -4 x y,
        \label{eq:tetra}
        \\
        \dot{x} = 2 x^2-48 y^2,
        \quad
        \dot{y}= -6 x y-8y^2,
        \label{eq:octa}
        \\
        \dot{x} = 2 x^2-y^2, 
        \quad
        \dot{y} = -10 x y +y^2.
        \label{eq:icos}
    \end{gather}
    \label{eq:rnahm}
\end{subequations}
Due to the symmetry of the associated Nahm matrices the
systems \eqref{eq:rnahm} are called the 
\emph{tetrahedral Nahm system}, 
\emph{octahedral Nahm system}, 
and \emph{icosahedral Nahm system} respectively.
The peculiarity of these systems is the fact that they
are \emph{algebraically integrable}, in the sense that they
possess an invariant elliptic curve, i.e. a genus one curve, 
of degree three, four and six respectively.
For more information on the general Nahm equations in the
context of the modern theory of integrable systems we 
refer to \cite{Ablowitz1991}.

In recent years arose the interest in the problem of finding good 
discretisation of continuous systems.
By good discretisation, here we mean a discretisation, which
preserves as much as possible the properties of its continuous
counterpart.
Within this framework a procedure called 
\emph{Kahan-Hirota-Kimura (KHK) discretisation} became popular as a way of 
producing integrable discrete equations from systems of integrables ODEs.
Specifically, given a systems of first-order ordinary differential equations:
\begin{equation}
    \vec{\dot{x}}=\vec{F}\left( \vec{x} \right)
    \label{eq:firstord}
\end{equation}
its KHK is given by the following formula:
\begin{equation}
    \frac{\vec{x}_{n+1}-\vec{x}_{n}}{h} = 
    2\vec{F}\left( \frac{\vec{x}_{n+1}+\vec{x}_{n}}{2} \right)
    -\frac{\vec{F}\left( \vec{x}_{n+1} \right)+\vec{F}\left( \vec{x}_{n}\right)}{2},
    \quad
    \vec{x}_{n} = \vec{x}\left( n h \right),\, h\to0^{+}.
    \label{eq:kahan}
\end{equation}
This formula was presented first by W. Kahan in a series of unpublished
lecture notes \cite{Kahan1993}, and applied by
K. Kimura and R. Hirota to produced an integrable discretisation of 
the Lagrange top \cite{HirotaKimura2000}, 
This result attracted the interest of many scientists working
in the field of geometric discretisation theory \cite{KrantzParks2008}, 
from the Berlin school
\cite{PetreraPfadlerSuris2009,PetreraSuris2010,PetreraPfadlerSuris2011}.
In particular in \cite{PetreraPfadlerSuris2009} it was noticed
that when the function $\vec{F}$ in \eqref{eq:firstord} is
quadratic the discretisation rule \eqref{eq:kahan} give raise
to a birational map.
Later, some general integrability properties of the 
KHK discretisation were unveiled through the
work of G. R.W. Quispel and his collaborators
\cite{CelledoniMcLachlanOwrenQuispel2013,CelledoniMcLachlanOwrenQuispel2014}.

In particular in \cite{PetreraPfadlerSuris2009,PetreraSuris2010,PetreraPfadlerSuris2011},
Petrera, Pfadler and Suris developed an algebraic approach for the
search of invariants for KHK discretisations, called the
\emph{Hirota--Kimura bases}.
With this approach they produced lots of examples, 
yet besides invariants and preserved measures,
little was know about additional structures of the
discrete integrable systems they found.
For instance, in the conclusions of \cite{PetreraPfadlerSuris2011}
the authors write:
\begin{quote}    
    ``Of course, it would be highly desirable to find some structures, 
    like Lax representation, bi-Hamiltonian structure, etc., which would allow 
    one to check the conservation of integrals in a more clever way, 
    but up to now no such structures have been found for any of the 
    [K]HK type discretizations.''
\end{quote}

In this paper, we give an answer to the above comment
made by Petrera, Pfadler and Suris in \cite{PetreraPfadlerSuris2011}.
That is, using a technique presented in \cite{Kimura2017Lax,Sogo2017},
we build the discrete analog of the reduced Nahm system
from their Lax representation.
Then, we show that this discretisation is equivalent to the
KHK discretisation discussed in \cite{PetreraPfadlerSuris2011}.
These Lax pairs are used to produce invariants, 
and proving integrability of the discrete Nahm systems.

The plan of the paper is the following: in section \ref{sec:laxcont}
we give a review of the literature on the Lax pair for the continuous 
and discrete Nahm systems.
In section \ref{sec:laxreddiscr} we use such construction to produce
the Lax pairs for the reduced Nahm systems \eqref{eq:rnahm} and prove
integrability.
In the final section \ref{sec:concl} we give some conclusions and an
outlook for further researches. 
Moreover, we show with an example that there exists Nahm systems
whose Lax pair does not provide integrability, 
yet the system is KHK discretisable, and both the continuous
and discrete systems are algebraically integrable.
This shows, that despite the success obtained in explaining the 
integrability of the Euler top \cite{Kimura2017Lax,Sogo2017,Kimura2017JPhysA},
and of the reduced Nahm systems \eqref{eq:rnahm},
the Nahm equation approach cannot solve the general problem of 
characterisation of the integrability of KHK discretisations.

\section{Lax pair for the continuous and discrete Nahm systems}
\label{sec:laxcont}

In the literature several different forms of the Lax pair
for the Nahm equations have been proposed.
For instance, recently in \cite{Kimura2017Lax} it was
proposed the following form:
\begin{subequations}
    \begin{align}
        A\left( \lambda \right) &=
        \begin{pmatrix}
            T_{1} & -T_{2}
            \\
            T_{2} & T_{1}
        \end{pmatrix}
        + \lambda 
        \begin{pmatrix}
            0 & 2T_{3}
            \\
            -2T_{3} & 0
        \end{pmatrix}
        + \lambda^{2}
        \begin{pmatrix}
            T_{1} & T_{2}
            \\
            -T_{2} & T_{1}
        \end{pmatrix}
        ,
        \label{eq:Akimura}
        \\
        B \left( \lambda \right) &=
        \begin{pmatrix}
            0 & T_{3}
            \\
            -T_{3} & 0
        \end{pmatrix}
        + \lambda
        \begin{pmatrix}
            T_{1} & T_{2}
            \\
            -T_{2} & T_{1}
        \end{pmatrix}
        \label{eq:Bkimura}
    \end{align}
    \label{eq:LPkimura}
\end{subequations}
The above matrices are such that the system \eqref{eq:nahm} is 
equivalent to the following compatibility condition:
\begin{equation}
    \dot{A} = \left[ A,B \right].
    \label{eq:LPcomp}
\end{equation}

The Lax pair \eqref{eq:LPkimura} consists of $2N\times 2N$ matrices.
In this paper to avoid too cumbersome formulas we consider
the inverse matrix complexifcation of the matrices in \eqref{eq:LPkimura}:
\begin{equation}
    M =\begin{pmatrix}
        M_{1} & -M_{2}
        \\
        M_{2} & M_{1}
    \end{pmatrix}
    \implies
    M = M_{1} + \imath M_{2}.
    \label{eq:decompl}
\end{equation}
That is, we consider the following
Lax pair for the Nahm system \eqref{eq:nahm}:
\begin{subequations}
    \begin{align}
        A\left( \lambda \right) &= T_{1} +\imath T_{2} 
         -2\imath\lambda T_{3}+ \lambda^{2} \left( T_{1} -\imath T_{2} \right),
        \label{eq:Anahm}
        \\
        B \left( \lambda \right) &= -\imath T_{3} +\lambda \left( T_{1} -\imath T_{2} \right).
        \label{eq:Bnahm}
    \end{align}
    \label{eq:LPnahm}
\end{subequations}
The compatibility condition \eqref{eq:LPcomp} gives again
the Nahm equations \eqref{eq:nahm} taking its real and 
imaginary part.

Associated to the matrix $B\left( \lambda \right)$
there exists a unique family of unitary matrices resolving the 
differential equation $\dot V = V B$, with initial condition 
$V\left( 0 \right)=I_{N}$.
As proven in \cite{Lax1968} this implies that the 
spectrum of the matrix $A\left( \lambda \right)$ does not depend
on the independent variable $t$.
So, the coefficients of the characteristic polynomial
of $A\left( \lambda \right)$:
\begin{equation}
    p_{A} \left( \mu \right) = \det \left(  A\left( \lambda \right)-\mu I_{N}  \right),
    \label{eq:detpol}
\end{equation}
do not depend on $t$.
That is, the coefficients of the characteristic polynomial \eqref{eq:detpol}
are first integrals of the Nahm system \eqref{eq:nahm}.
Moreover, since the system \eqref{eq:nahm} does not depend on
the variable $\lambda$ too, 
for each coefficient we can have multiple first integrals.

\begin{remark}
    Equation \eqref{eq:detpol} yields at least $N$ first integrals,
    but there is no \emph{a priori} guarantee that these 
    first integrals are functionally independent and/or non-trivial.
    This implies that the integrability of the system \eqref{eq:nahm}
    must be proved case by case using the appropriate form of the
    matrices $T_{i}$.
    \label{rem:indip}
\end{remark}

In \cite{Kimura2017JPhysA}, followed by \cite{Kimura2017Lax,Sogo2017}, was 
introduced a method to discretise the compatibility condition \eqref{eq:LPcomp}.
Consider the discrete time interval:
\begin{equation}
    t_{n} = n h, \quad h\to 0^{+},
    \label{eq:tn}
\end{equation}
hence we define $f_{n}\equiv f(t_{n})$.
Then, the compatibility condition \eqref{eq:LPcomp} can be
discretised as:
\begin{equation}
    \frac{A_{n+1}\left( \lambda \right)-A_{n}\left( \lambda \right)}{h} 
    = A_{n+1}\left( \lambda \right)B_{n}\left( \lambda \right)
    -B_{n+1}\left( \lambda \right)A_{n}\left( \lambda \right).
    \label{eq:LPdiscr}
\end{equation}
The corresponding system of difference equations is given by:
\begin{equation}
    \frac{T_{i,n+1}-T_{i,n}}{h}
    = 
    T_{j,n+1}T_{k,n}-T_{k,n+1}T_{j,n},
    \label{eq:dnahm}
\end{equation}
where the indices $i,j,k$ are cyclic permutations of the set
$\left\{ 1,2,3 \right\}$.

\begin{remark}
    In principle a different discretisation of the compatibility
    condition \eqref{eq:LPcomp} can be given:
    \begin{equation}
        \frac{A_{n+1}\left( \lambda \right)-A_{n}\left( \lambda \right)}{h} 
        = A_{n}\left( \lambda \right)B_{n+1}\left( \lambda \right)
        -B_{n}\left( \lambda \right)A_{n+1}\left( \lambda \right),
        \label{eq:LPdiscrinv}
    \end{equation}
    yield the following system of difference equations:
    \begin{equation}
        \frac{T_{i,n+1}-T_{i,n}}{h}
        = 
        T_{j,n}T_{k,n+1}-T_{k,n}T_{j,n+1},
        \label{eq:dnahminv}
    \end{equation}
    where the indices $i,j,k$ are cyclic permutations of the set
    $\left\{ 1,2,3 \right\}$.
    However, by direct computation it is possible to show that,
    in the cases considered in this paper,
    condition \eqref{eq:dnahminv} is equivalent to 
    \eqref{eq:dnahm} up to the transformation:
    \begin{equation}
        \vec{x}_{n+i} \longleftrightarrow \vec{x}_{n-i},
        \label{eq:equiv}
    \end{equation}
    where $\vec{x}_{n}$ is the vector of the dynamical variables.
    That is, the evolution defined from \eqref{eq:dnahm} is the
    opposite of the evolution defined by \eqref{eq:dnahminv}.
    We notice that this is a general fact when dealing with 
    KHK discretisation as pointed out in \cite{PetreraPfadlerSuris2009}.
    \label{rem:inversediscr}
\end{remark}

Equation \eqref{eq:LPdiscr} can be rearranged as:
\begin{equation}
    A_{n+1}\left( \lambda \right) \left( I_{N}-hB_{n}\left( \lambda \right) \right) 
    = 
    \left( I_{N}-hB_{n+1}\left( \lambda \right) \right)A_{n}\left( \lambda \right).
    \label{eq:LPdiscr2}
\end{equation}
Introducing the matrices:
\begin{equation}
    L_{n}\left( \lambda \right) = A_{n}\left( \lambda \right), 
    \quad 
    M_{n}\left( \lambda \right) = I_{N}-hB_{n}\left( \lambda \right),
    \label{eq:LPdiscrdef}
\end{equation}
which allows us to rewrite \eqref{eq:LPdiscr2} as:
\begin{equation}
    L_{n+1}\left( \lambda \right)M_{n}\left( \lambda \right) 
    = 
    M_{n+1}\left( \lambda \right)L_{n}\left( \lambda \right).
    \label{eq:LPdiscr3}
\end{equation}
Following \cite{Suris2003book,Suris1990} we have that equation
\eqref{eq:LPdiscr3} implies that the spectral data of the matrix
$L_{n}\left( \lambda \right)M_{n}^{-1}\left( \lambda \right)$ are constant
along the evolution.
Indeed, from \eqref{eq:LPdiscr3}, the matrices 
$L_{n+1}\left( \lambda \right) M_{n+1}^{-1}\left( \lambda \right)$ and
$L_{n}\left( \lambda \right)M_{n}^{-1}\left( \lambda \right)$ are conjugate,
so that they have the same characteristic polynomial.
This implies that the that the coefficients of the 
characteristic polynomial of $L_{n}\left( \lambda \right)M_{n}^{-1}\left( \lambda \right)$ 
are conserved quantities (invariants) for the system \eqref{eq:dnahm}. 
Alternatively, using Binet's rule, we have 
that the coefficients of the 
characteristic polynomial of $L_{n}\left( \lambda \right)$ with respect
to $M_{n}\left( \lambda \right)$:
\begin{equation}
    p_{L,M} \left( \mu \right) = 
    \det \left(  L_{n}\left( \lambda \right)-\mu M_{n}\left( \lambda \right)  \right),
    \label{eq:discrdetpol}
\end{equation}
divided by $\det M_{n}\left( \lambda \right)$ are constants of motions.
That is, writing such characteristic polynomial in the following way:
\begin{equation}
    p_{L,M} \left( \mu \right) = 
    \left( -1 \right)^{N}\det M_{n}\left( \lambda \right)\mu^{N}+
    c_{N-1}\left( \lambda \right)\mu^{N-1}+
    \dots
    +c_{0}\left( \lambda \right),
    \label{eq:discrdetpol2}
\end{equation}
we can write these invariants in the following way:
\begin{equation}
    H_{0} = \frac{c_{0}\left( \lambda \right)}{\det M_{n}\left( \lambda \right)},
    \,
    H_{1} = \frac{c_{1}\left( \lambda \right)}{\det M_{n}\left( \lambda \right)},
    \,
    \dots,
    \,
    H_{N-1} = \frac{c_{N-1}\left( \lambda \right)}{\det M_{n}\left( \lambda \right)}.
    \label{eq:H0N}
\end{equation}
Finally, we note that the same consideration on functional independence 
of the invariants \eqref{eq:H0N} given in Remark \ref{rem:indip} apply.

\section{Discrete reduced Nahm systems}
\label{sec:laxreddiscr}

In \cite{Hitchinetal1995} where considered three special cases of Nahm's
equations \eqref{eq:nahm} corresponding to symmetry groups of regular
solids, namely \emph{tetrahedral}, \emph{octahedral}, 
and \emph{icosahedral} symmetry.

Assume we are a given $G\subset SO\left( 3 \right)$, 
a symmetry group of regular solid.
Then, the $G$-invariant Nahm matrices $T_{i}$ have
the following form:
\begin{equation}
    T_{i}\left( t \right) = x\left( t \right) \rho_{i} + y\left( t \right) S_{i}.
    \label{eq:hitchint}
\end{equation}
Here $\rho\colon\R^{3}\to\su\left( k \right)$ is 
a representation of $\so\left( 3 \right)$ on $\Cp^{k}$,
while $\left( S_{1},S_{2},S_{3} \right)$ is a $G$-invariant vector in the
symmetric power space $S^{2k}V\subset \R^{3}\otimes\su\left( k \right)$ 
where $V$ is the representation corresponding to $G$ in $SU\left( 2 \right)$.

In the following we will consider the tetrahedral, octahedral, 
and icosahedral symmetry cases, with the definitions of the 
$G$-invariant Nahm matrices $T_{i}$ given in \cite{Hitchinetal1995}.
The discretisation of these continuous systems was obtained
in \cite{PetreraPfadlerSuris2011} using the KHK discretisation procedure
and proved to be integrable by constructing the invariant with
the so-called Hirota--Kimura bases \cite{PetreraPfadlerSuris2009}.
Here we will prove that the KHK discretisation follows from the
discretisation of the Lax pairs and the invariant can be found using
the associated characteristic polynomial \eqref{eq:discrdetpol}.

\begin{remark}
    We note that the results of \cite{PetreraPfadlerSuris2009} on
    the discrete Nahm systems where generalised simultaneously and
    independently in 
    \cite{CelledoniMcLachlanMcLarenOwrenQuispel2017,PetreraZander2017}.
    Some comments on the geometry of these systems were given
    in \cite{CarsteaTakenawa2013}.
    Later, in \cite{VanDerKampCelledoniMcLachlanMcLarenOwrenQuispel2019}
    it was proved how to construct the tetrahedral and the octahedral case
    discrete cases using generalised Manin transform.
    Finally, in \cite{GJ_biquadratic} it was pointed out that in the octahedral
    case the geometric
    construction of \cite{VanDerKampCelledoniMcLachlanMcLarenOwrenQuispel2019}
    induces non-standard features on the procedure of resolution of
    singularities, proving the existence of families of particular solutions.
    \label{rem:gen}
\end{remark}

\subsection{Tetrahedral symmetry}
\label{sec:tetra}

Consider the reduced Nahm system with tetrahedral symmetry \eqref{eq:tetra}.
Its Lax pair is given by:
\begin{subequations}
    \begin{align}
        A\left( \lambda \right) &=
        \begin{pmatrix}
            0&2 \imath\lambda  \left( 2 x +\frac{y}{2}  \right) &-\imath\left({\lambda}^{2}-1\right) \left( 2 x- \frac{y}{2}  \right) 
            \\ 
            -2 \imath\lambda  \left( 2 x -\frac{y}{2} \right) &0&-\left({\lambda}^{2}+1\right)\left( 2 x +\frac{y}{2}  \right) 
            \\ 
            \imath\left({\lambda}^{2}-1\right) \left( 2 x+ \frac{y}{2}  \right)  &
            \left({\lambda}^{2}+1\right)\left( 2 x -\frac{ y}{2}  \right) &0
        \end{pmatrix},
        \label{eq:tetraL}
        \\
        B\left( \lambda \right) &=
        \begin{pmatrix}
            0&\imath \left( 2 x + \frac{y}{2}  \right) &-\imath \lambda  \left( 2 x - \frac{y}{2} \right) 
            \\ 
            -\imath \left( 2 x -\frac{y}{2}  \right) &0& -\lambda  \left( 2 x +\frac{y}{2} \right) 
            \\
            \imath\lambda  \left( 2 x +\frac{y}{2} \right) &\lambda  \left( 2 x -\frac{y}{2}  \right) &0
        \end{pmatrix}.
        \label{eq:tetraM}
    \end{align}
    \label{eq:tetraLM}
\end{subequations}
Considering the characteristic polynomial \eqref{eq:detpol}
we obtain the invariant given in \cite{Hitchinetal1995}:
\begin{equation}
    H = y (y^2+48x^2).
    \label{eq:tetraH}
\end{equation}
Note that the level curves of the invariant $H$ \eqref{eq:tetraH} 
are genus one (elliptic) curves.

From \eqref{eq:LPdiscrdef} we obtain the following
discrete Lax Pair:
\begin{subequations}
    \begin{align}
        L_{n}\left( \lambda \right) &=
        \begin{pmatrix}
            0&2 \imath\lambda  \left( 2 x_{n} +\frac{y_{n}}{2}  \right) &-\imath\left({\lambda}^{2}-1\right) \left( 2 x_{n}- \frac{y_{n}}{2}  \right) 
            \\ 
            -2 \imath\lambda  \left( 2 x_{n} -\frac{y_{n}}{2} \right) &0&-\left({\lambda}^{2}+1\right)\left( 2 x_{n} +\frac{y_{n}}{2}  \right) 
            \\ 
            \imath\left({\lambda}^{2}-1\right) \left( 2 x_{n}+ \frac{y_{n}}{2}  \right)  &
            \left({\lambda}^{2}+1\right)\left( 2 x_{n} -\frac{y_{n}}{2}  \right) &0
        \end{pmatrix},
        \label{eq:tetraLd}
        \\
        M_{n}\left( \lambda \right) &=
        \begin{pmatrix}
            1&-\imath h \left( 2 x_{n} +\frac{y_{n}}{2}  \right) &\imath \lambda h  \left( 2 x_{n} - \frac{y_{n}}{2} \right) 
            \\ 
            \imath h \left( 2 x_{n} -\frac{y_{n}}{2}  \right) &1& \lambda h \left( 2 x_{n} +\frac{y_{n}}{2} \right) 
            \\
            -\imath\lambda h \left( 2 x_{n} +\frac{y_{n}}{2} \right) &-\lambda h \left( 2 x_{n} -\frac{y_{n}}{2}  \right) &1
        \end{pmatrix}.
        \label{eq:tetraMd}
    \end{align}
    \label{eq:tetraLMd}
\end{subequations}
The corresponding compatibility conditions are:
\begin{subequations}
    \begin{align}
        \frac{x_{n+1}-x_{n}}{h} &= 2 x_{n+1} x_{n}+\frac{1}{8} y_{n+1} y_{n},
        \\
        \frac{y_{n+1}-y_{n}}{h} &= -2 \left(x_{n} y_{n+1}+x_{n+1} y_{n}\right),
    \end{align}
    \label{eq:tetrad}
\end{subequations}
and coincide with the KHK discretisation of the reduced
Nahm system with tetrahedral symmetry \eqref{eq:tetra},
originally presented in \cite{PetreraPfadlerSuris2011}.
The characteristic polynomial of \eqref{eq:tetraLd} with
respect to \eqref{eq:tetraMd}:
\begin{equation}
    \begin{aligned}
        p_{L,A} \left( \mu \right)&=
    \left(1-4 h^2 x_{n}^2+ \frac{h^2 y_{n}^2}{4}+12 h^3 x_{n}^2 \lambda^2 y_{n}+ \frac{h^3 y_{n}^3}{4} \lambda^2\right) \mu^3
        \\
        &-h^2 \lambda^3 y_{n} (48 x_{n}^2+y_{n}^2) \mu^2
        + \frac{h y_{n}}{4} (48 x_{n}^2+y_{n}^2) (5 \lambda^4-1) \mu
        \\
        &-\frac{\lambda y_{n}}{2} (48 x_{n}^2+y_{n}^2) (\lambda-1) (\lambda+1) (\lambda^2+1).
    \end{aligned}
    \label{eq:ptetra}
\end{equation}
From equation \eqref{eq:H0N} we could get three different invariants,
yet they will be necessarily dependent.
So, we choose:
\begin{equation}
    H_{0}\left( \lambda \right) =
    -\frac{1}{2} \frac{\lambda y_{n}(48 x_{n}^2+y_{n}^2) (\lambda-1) (\lambda+1) (\lambda^2+1)}{
        \displaystyle
    1-4 h^2 x_{n}^2+ \frac{h^2 y_{n}^2}{4}+12 h^3 x_{n}^2 \lambda^2 y_{n}+ \frac{h^3 y_{n}^3}{4} \lambda^2}.
    \label{eq:H0tetra}
\end{equation}
This invariant is dependent on $\lambda$, therefore to find a
non-trivial invariant we can expand in Taylor series with respect to $\lambda$,
that is
$H_{0}\left( \lambda \right) = \sum_{k=0}^{\infty}h_{0,k}\lambda^{k}$,
and take the first non-constant element:
\begin{equation}
    h_{0,1} =
    -\frac{1}{2} \frac{ y_{n}(48 x_{n}^2+y_{n}^2)}{
    1-4 h^2 x_{n}^2+ \displaystyle\frac{h^2 y_{n}^2}{4}}.
    \label{eq:H0tetra2}
\end{equation}
This is an invariant for \eqref{eq:tetrad}.

\subsection{Octahedral symmetry}
Consider the reduced Nahm system with tetrahedral symmetry \eqref{eq:octa}.
%We remark that this system is linearly equivalent to a system
%previously considered in 
%\cite{GJ_biquadratic,CelledoniMcLachlanMcLarenOwrenQuispel2017}.
Its Lax pair is given by:
\begin{subequations}
    \begin{align}
        A\left( \lambda \right) &=
        \begin{pmatrix}
            2 \lambda  \left( 3 x+4 y \right) &6 {\lambda}^{2} \left( x -2y\right) &0&-120 y  
            \\ 
            -2 x+4 y   &2 \lambda  \left( x   -12 y  \right) &8 {\lambda}^{2} \left(x+3 y\right) &0
            \\ 
            0&-2 x -6 y   &-2 \lambda\left( x -12 y    \right) &6 {\lambda}^{2} \left( x   -2 y    \right) 
            \\ 
            10 {\lambda}^{2}y/3   &0&-2 x   +4 y   &-2 \lambda  \left( 3 x+4 y \right)            
        \end{pmatrix},
        \label{eq:octaL}
        \\
        B\left( \lambda \right) &=
        \begin{pmatrix}
            3x +4 y   &6 \lambda  \left( x-2 y\right) &0&0
            \\ 
            0&x   -12 y   &8 \lambda  \left( x   +3 y    \right) &0
            \\ 
            0&0&-\left(x   -12 y\right)   &6 \lambda  \left( x -2 y\right) 
            \\
            10\lambda y/3   &0&0&-\left(3 x   +4 y\right)
        \end{pmatrix}.
        \label{eq:octaM}
    \end{align}
    \label{eq:octaLM}
\end{subequations}
Considering the characteristic polynomial \eqref{eq:detpol}
we obtain the invariant given in \cite{Hitchinetal1995}:
\begin{equation}
    H = y(x+3y)(x-2y)^2.
    \label{eq:octaH}
\end{equation}
Note that the level curves of the invariant $H$ \eqref{eq:octaH} 
are genus one (elliptic) curves.

From \eqref{eq:LPdiscrdef} we obtain the following
discrete Lax Pair:
\begin{subequations}
    \begin{align}
        L_{n}\left( \lambda \right) &=
        \begin{pmatrix}
            2 \lambda  \left( 3 x_{n}+4 y_{n} \right) &6 {\lambda}^{2} \left( x_{n} -2y_{n}\right) &0&-120 y_{n}  
            \\ 
            -2 x_{n}+4 y_{n}   &2 \lambda  \left( x_{n}   -12 y_{n}  \right) &8 {\lambda}^{2} \left(x_{n}+3 y_{n}\right) &0
            \\ 
            0&-2 x_{n} -6 y_{n}   &-2 \lambda\left( x_{n} -12 y_{n}    \right) &6 {\lambda}^{2} \left( x_{n}   -2 y_{n}    \right) 
            \\ 
            10 {\lambda}^{2}y_{n}/3   &0&-2 x_{n}   +4 y_{n}   &-2 \lambda  \left( 3 x_{n}+4 y_{n} \right)            
        \end{pmatrix},
        \label{eq:octaLd}
        \\
        M_{n}\left( \lambda \right) &=
        \begin{pmatrix}
            1-h\left(3x_{n} +4 y_{n}\right)   &1- 6 \lambda h  \left( x_{n}-2 y_{n}\right) &0&0
            \\ 
            0&1-h\left(x_{n}   -12 y_{n}\right) &1- 8 \lambda h \left( x_{n}   +3 y_{n}    \right) &0
            \\ 
            0&0&1+h\left(x_{n}   -12 y_{n}\right)   &1- 6 h \lambda  \left( x_{n} -2 y_{n}\right) 
            \\
            -10\lambda h y_{n}/3   &0&0&1+h\left(3 x_{n}+4 y_{n}\right)
        \end{pmatrix}.
        \label{eq:octaMd}
    \end{align}
    \label{eq:octaLMd}
\end{subequations}
The corresponding compatibility conditions are:
\begin{subequations}
    \begin{align}
        {\frac {x_{{n+1}}-x_{{n}}}{h}} &=2 x_{{n+1}}x_{{n}}-48 y_{{n+1}}y_{{n}},  
        \\
        \frac {y_{{n+1}}-y_{{n}}}{h} &=-3 \left(x_{{n+1}}y_{{n}}+ x_{{n}}y_{n+1}\right)
        -8 y_{n} y_{n+1}.
    \end{align}
    \label{eq:octad}
\end{subequations}
and coincide with the KHK discretisation of the reduced
Nahm system with octahedral symmetry \eqref{eq:octa}, 
originally presented in \cite{PetreraPfadlerSuris2011}.
Proceeding in analogous ways as in section \ref{sec:tetra},
that is taking the first non-constant element of the Taylor series of
$H_{0}\left( \lambda \right)$, we obtain the following
invariant:
\begin{equation}
    h_{0,0} =
    {\frac {-960y_{{n}} \left( x_{{n}}+3 y_{{n}} \right)  \left( x_{{n}}-2 y_{{n}} \right) ^{2}}{%
            \left[1- h^{2}\left(x_{{n}}-12y_{{n}}\right)^{2} \right]  
    \left[ 1-h^{2}  \left(3 x_{{n}}+4 y_{{n}}\right)^{2} \right] }}.
    \label{eq:H0octa}
\end{equation}

\subsection{Icosahedral symmetry}
Consider the reduced Nahm system with tetrahedral symmetry \eqref{eq:icos}.
%The reduced Nahm equation with icosahedral symmetry is:
%\begin{equation}
%    \dot{x} = 2 x^2-z^2, 
%    \quad
%    \dot{z} = -10 x z +z^2.
%    \label{eq:icos}
%\end{equation}
%We remark that this system is linearly equivalent to the
%sextic example considered in 
%\cite{CelledoniMcLachlanMcLarenOwrenQuispel2017}.
Its Lax pair is given by:
\begin{subequations}
    \begin{align}
        L\left( \lambda \right) &=
        \begin{pmatrix}
                \frac{2\lambda}{5} \left( 25 x +y \right) &
                {\lambda}^{2} \left( 10 x -y  \right) &0&0&-168 y &336 \lambda y
                \\ 
                -2 x + \frac{y}{5} &2 \lambda  \left( 3 x -y  \right) &
                4 {\lambda}^{2}\left( 4 x +y  \right) &0&0&168 y
                \\ 
                0&-2 x -\frac{y}{2} &2 \lambda  \left( x +2 y \right) & 
                6 {\lambda}^{2} \left( 3 x -y  \right) &0&0
                \\
                0&0&-2x +2/3 y &-2 \lambda  \left( x +2 y  \right) &4 {\lambda}^{2}
         \left( 4 x +y  \right) &0
            \\ 
            {\frac {7 {\lambda}^{2}y}{120}}&0&0&-2 x -\frac{y}{2} &
            -2 \lambda \left( 3 x -y \right) &{\lambda}^{2} \left( 10 x -y \right) 
            \\ 
            {\frac {7 \lambda y}{300}}&-{\frac {7 {\lambda}^{2}y}{120}}&0&0&-2 x+\frac{y}{5} &-\frac{2}{5}\lambda  \left( 25 x +y  \right) 
        \end{pmatrix},
        \label{eq:icosL}
        \\
        M\left( \lambda \right) &=
        \begin{pmatrix}
            5 x +\frac{y}{5} &\lambda  \left( 10 x -y\right) &0&0&0&168 y 
            \\ 
            0&3 x -y &4 \lambda  \left( 4 x +y   \right) &0&0&0
            \\ 
            0&0&x +2 y &6 \lambda  \left( 3 x -y  \right) &0&0
            \\ 
            0&0&0&-x -2 y &4 \lambda  \left( 4 x +y  \right) &0
            \\ 
            {\frac {7 \lambda y}{120}}&0&0&0&-3 x +y & \lambda  \left( 10 x -y  \right)
            \\ 
            {\frac {7 y}{600}}&-{\frac {7 \lambda y }{120}}&0&0&0&-5 x - \frac{y}{5}
        \end{pmatrix}.
        \label{eq:icosM}
    \end{align}
    \label{eq:icosLM}
\end{subequations}
Considering the characteristic polynomial \eqref{eq:detpol}
we obtain the invariant given in \cite{Hitchinetal1995}:
\begin{equation}
    H = y(3x-y)^2(4x+y)^3.
    \label{eq:icosH}
\end{equation}
Note that the level curves of the invariant $H$ \eqref{eq:icosH} 
are genus one (elliptic) curves.

\begin{landscape}
From \eqref{eq:LPdiscrdef} we obtain the following
discrete Lax Pair:
\begin{subequations}
%        \begin{equation}
%        L_{n}\left( \lambda \right) &=
%        \begin{pmatrix}
%                \frac{2\lambda}{5} \left( 25 x_{n} +y_{n} \right) &
%                {\lambda}^{2} \left( 10 x_{n} -y_{n}  \right) &0&0&-168 y_{n} &336 \lambda y_{n}
%                \\ 
%                -2 x_{n} + \frac{y_{n}}{5} &2 \lambda  \left( 3 x_{n} -y_{n}  \right) &
%                4 {\lambda}^{2}\left( 4 x_{n} +y_{n}  \right) &0&0&168 y_{n}
%                \\ 
%                0&-2 x_{n} -\frac{y_{n}}{2} &2 \lambda  \left( x_{n} +2 y_{n} \right) & 
%                6 {\lambda}^{2} \left( 3 x_{n} -y_{n}  \right) &0&0
%                \\
%                0&0&-2x_{n} +2/3 y_{n} &-2 \lambda  \left( x_{n} +2 y_{n}  \right) &4 {\lambda}^{2}
%         \left( 4 x_{n} +y_{n}  \right) &0
%            \\ 
%            {\frac {7 {\lambda}^{2}y_{n}}{120}}&0&0&-2 x_{n} -\frac{y_{n}}{2} &
%            -2 \lambda \left( 3 x_{n} -y_{n} \right) &{\lambda}^{2} \left( 10 x_{n} -y_{n} \right) 
%            \\ 
%            {\frac {7 \lambda y_{n}}{300}}&-{\frac {7 {\lambda}^{2}y_{n}}{120}}&0&0&-2 x_{n}+\frac{y_{n}}{5} &-\frac{2}{5}\lambda  \left( 25 x_{n} +y_{n}  \right) 
%        \end{pmatrix},
%        \label{eq:icosLd}
%        \end{equation}
    \begin{align}
        L_{n}\left( \lambda \right) &=
        \begin{pmatrix}
                \frac{2\lambda}{5} \left( 25 x_{n} +y_{n} \right) &
                {\lambda}^{2} \left( 10 x_{n} -y_{n}  \right) &0&0&-168 y_{n} &336 \lambda y_{n}
                \\ 
                -2 x_{n} + \frac{y_{n}}{5} &2 \lambda  \left( 3 x_{n} -y_{n}  \right) &
                4 {\lambda}^{2}\left( 4 x_{n} +y_{n}  \right) &0&0&168 y_{n}
                \\ 
                0&-2 x_{n} -\frac{y_{n}}{2} &2 \lambda  \left( x_{n} +2 y_{n} \right) & 
                6 {\lambda}^{2} \left( 3 x_{n} -y_{n}  \right) &0&0
                \\
                0&0&-2x_{n} +2/3 y_{n} &-2 \lambda  \left( x_{n} +2 y_{n}  \right) &4 {\lambda}^{2}
         \left( 4 x_{n} +y_{n}  \right) &0
            \\ 
            {\frac {7 {\lambda}^{2}y_{n}}{120}}&0&0&-2 x_{n} -\frac{y_{n}}{2} &
            -2 \lambda \left( 3 x_{n} -y_{n} \right) &{\lambda}^{2} \left( 10 x_{n} -y_{n} \right) 
            \\ 
            {\frac {7 \lambda y_{n}}{300}}&-{\frac {7 {\lambda}^{2}y_{n}}{120}}&0&0&-2 x_{n}+\frac{y_{n}}{5} &-\frac{2}{5}\lambda  \left( 25 x_{n} +y_{n}  \right) 
        \end{pmatrix},
        \label{eq:icosLd}
        \\
        M_{n}\left( \lambda \right) &=
        \begin{pmatrix}
            1-h\left(5 x_{n} +\frac{y_{n}}{5}\right) &-\lambda h \left( 10 x_{n} -y_{n}\right) &0&0&0& -168 h y_{n} 
            \\ 
            0& 1-h\left(3 x_{n} -y_{n}\right) &-4h \lambda  \left( 4 x_{n} +y_{n}   \right) &0&0&0
            \\ 
            0&0&1-h\left(x_{n} +2 y_{n}\right) &-6h \lambda  \left( 3 x_{n} -y_{n}  \right) &0&0
            \\ 
            0&0&0&1+h\left(x_{n} +2 y_{n}\right) &-4 h\lambda  \left( 4 x_{n} +y_{n}  \right) &0
            \\ 
            -{\frac {7h \lambda y_{n}}{120}}&0&0&0&1+h\left(3 x_{n} -y_{n}\right) & -\lambda h \left( 10 x_{n} -y_{n}  \right)
            \\ 
            -{\frac {7h y_{n}}{600}}&{\frac {7h \lambda y_{n} }{120}}&0&0&0&1+h\left(5 x_{n} + \frac{y_{n}}{5}\right)
        \end{pmatrix}.
        \label{eq:icosMd}
    \end{align}
    \label{eq:icosLMd}
\end{subequations}
\end{landscape}
The corresponding compatibility conditions are:
\begin{subequations}
    \begin{align}
        {\frac {x_{n+1}-x_{n}}{h}} &= 2x_{n+1} x_{n} - y_{n+1} y_{n},
        \\
        {\frac {y_{n+1}-y_{n}}{h}} &=-5 \left(x_{n}y_{n+1}+x_{n+1} y_{n}\right) 
        + y_{n} y_{n+1}. 
    \end{align}
    \label{eq:icosd}
\end{subequations}
and coincide with the KHK discretisation of the reduced
Nahm system with icoshedral symmetry \eqref{eq:icos}, 
originally presented in \cite{PetreraPfadlerSuris2011}.
Proceeding in analogous ways as in section \ref{sec:tetra},
that is taking the first non-constant element of the Taylor series of
$H_{0}\left( \lambda \right)$, we obtain the following invariant:
\begin{equation}
    h_{0,1} ={\frac {112 y_{{n}} 
        \left( 3 x_{{n}}-y_{{n}} \right) ^{2} 
        \left( 4 x_{{n}}+y_{{n}} \right) ^{3}}{% 
            \left[1-{h}^{2}\left( 25 x_{n}^{2}+2 x_{n}y_{n}+2 y_{{n}}^{2}\right)\right]
            \left[ 1-h^{2}\left(x_{{n}}-2 y_{{n}}\right)^{2} \right]
            \left[ 1-h^{2}\left(3 x_{{n}}-y_{{n}}\right)^{2} \right] }}.
    \label{eq:H0icos}
\end{equation}

\section{Conclusions}
\label{sec:concl}

In this paper we discretised the reduced Nahm systems \cite{Hitchinetal1995} 
using the technique employed for the Euler top presented in 
\cite{Kimura2017Lax,Sogo2017,Kimura2017JPhysA}.
Our results shows that the discretisation is analog to the so-called 
Kahan--Hirota--Kimura discretisation.
We proved that such Lax pairs are ``bona fide''. 
That is, they can be used to produce (all) the invariants of the associated 
systems, and hence to prove integrability.
A Lax pair that cannot be used to produce (all) the invariants
of a given system is called a \emph{fake Lax pair}
\cite{CalogeroNucci1991,HayButler2013,HayButler2015}.

We conclude this paper noting that, unfortunately 
Nahm systems and their generalisations are not enough to
explain the integrability of all KHK discretisable systems.
To this end we consider the following system, which is a particular
case of the coupled Euler top introduced in \cite{Golseetal2008}:
\begin{subequations}
    \begin{align}
        \dot{x}_1 &=x_2 x_3,
        \label{eq:e1nd}
        \\
        \dot{x}_2 &= x_1 x_3 ,
        \label{eq:e2nd}
        \\
        \dot{x}_3 &= x_{1}x_{2}+ x_{4}x_5,
        \label{eq:e3nd}
        \\
        \dot{x}_4  &=x_3 x_5,
        \label{eq:e4nd}
        \\
        \dot{x}_5 &=x_3 x_4. 
        \label{eq:e5nd}
    \end{align}
    \label{eq:end}
\end{subequations}
This system is na\"ively integrable. 
We say that system of difference equations is na\"ively integrable 
when it possesses $N-1$ functionally independent first integrals
(invariants), where $N$ is the number of degrees of freedom.
In \cite{PetreraPfadlerSuris2011} it was proved that the system
\eqref{eq:end} possesses five first integrals, four of which are functionally
independent, proving na\"ive integrability.

It is easy to see that the system \eqref{eq:end} arises from the
following Nahm system:
\begin{subequations}
    \begin{align}
        T_{1} &=
        \begin{pmatrix}
            1&-x_1 &0
             \\ 
             x_1  &1& x_4
             \\ 
             0&-x_4 &1
         \end{pmatrix},
        \label{eq:T1nd}
        \\
        T_{2} &=
        \begin{pmatrix}
            1&0& x_3
            \\ 
            0&1&0
            \\
            -x_{3} &0&1
        \end{pmatrix},
        \label{eq:T2nd}
         \\
         T_{3} &=
         \begin{pmatrix}
             1& x_5 &0
             \\ 
             -x_5 & 1 & x_2
             \\ 
             0&-x_2 &1
         \end{pmatrix}.
        \label{eq:T3nd}
    \end{align}
    \label{eq:nahmnd}
\end{subequations}

So, the system \eqref{eq:end} has a Lax pair given by
\eqref{eq:LPnahm}:
\begin{subequations}
    \begin{align}
        A\left( \lambda \right) &=
        \begin{pmatrix}
            \left( 1-\imath \right) {\lambda}^{2}-2 \imath\lambda+1+i &
            -{\lambda}^{2}x_1  -2 \imath\lambda x_5 - x_1 &
            -\imath{\lambda}^{2} x_3 +\imath x_3 
            \\ 
            {\lambda}^{2} x_1 +2 \imath \lambda x_5  + x_1 &
            \left( 1-\imath \right) {\lambda}^{2}-2 \imath\lambda+1+i &
            {\lambda}^{2} x_4 -2 \imath \lambda x_2 + x_4 
            \\ 
            \imath{\lambda}^{2} x_3 -\imath x_3 &
            -{\lambda}^{2} x_4 +2 \imath\lambda x_2 - x_4 & 
            \left( 1-\imath \right) {\lambda}^{2}-2 \imath \lambda+1+\imath
        \end{pmatrix},
        \label{eq:ACET2}
        \\
        B\left( \lambda \right) &=
        \begin{pmatrix}
            -\imath+ \left( 1-\imath \right) \lambda&
            -\imath x_5 -\lambda x_1 &
            -\imath x_3 
            \\ 
            \imath x_5 +\lambda x_1 &
            -\imath+ \left( 1-\imath \right) \lambda&
            -\imath x_2 +\lambda x_4 
            \\ 
            \imath\lambda  x_3 &
            \imath x_2 -\lambda x_4 &
            -\imath+ \left( 1-\imath \right) \lambda
        \end{pmatrix},
        \label{eq:BCET2}
    \end{align}
    \label{eq:LaxCET2}
\end{subequations}
and the characteristic polynomial \eqref{eq:detpol} of $L\left( \lambda \right)$
is:
\begin{equation}
    \begin{aligned}
        p_{L}\left( \mu \right)
        &=
        {\mu}^{3} 
        + \left( 3-3 \imath \right)  \left( \imath\lambda-{\lambda}^{2}-\imath-\lambda \right) {\mu}^{2}
        \\
        &+
        \left[ 
            \begin{gathered}
                \left(  \I_{1} -6 \imath \right) {\lambda}^{4}
                +4 \imath \left( \I_{2} -3+3 \imath \right) {\lambda}^{3}
                -4 \I_{3} \lambda^{2}
                +4 \imath \left( \I_{2} -3 -3 \imath \right) \lambda
                + \I_{1} +6 \imath
            \end{gathered}
        \right]\mu
        \\
        &-\left( 1-\imath \right)  
        \left( \I_{1} - 2 \imath \right) {\lambda}^{6}
        +2 \imath \left[ 
            2 \left(\imath-1\right) \I_{2} + \I_{1}-6 \imath \right] {\lambda}^{5}
        \\
        &- \frac{3-\imath}{5} 
        \left[ 5 \I_1+(12+4\imath)\I_2 -(8-4\imath)\I_3-12+6\imath \right] \lambda^{4}
        \\
    &+4 \imath
    \left( 4+\I_{1}-2\I_{2}-2\I_{3} \right) {\lambda}^{3}
        \\
        &+ \frac{3+\imath}{5} \left( 12+6\imath -5\I_1-4(3-\imath) \I_2 +4(2+\imath)\I_3  \right) {\lambda}^{2}
    \\
    &-2 \imath\left[ 2 \left(1+\imath\right) \I_{2} -\I_{1} -6 i \right] \lambda 
        - \left( 1+i \right)  \left( \I_{1} +2 i \right)        
    \end{aligned}
    \label{eq:pCET2}
\end{equation}
where:
\begin{equation}
    \mathcal{I}_{1} = x_1^{2}- x_3^{2}+ x_4^{2},
    \quad
    \mathcal{I}_{2} = x_1 x_5  - x_4 x_2,
    \quad 
    \I_{3} = x_{2}^{2}-x_{3}^{2}+x_{5}^{2}.
    \label{eq:ndint}
\end{equation}
Taking the coefficients of \eqref{eq:pCET2} with respect to
$\lambda$ and $\mu$ we obtain that the three functions in \eqref{eq:ndint}
are the only independent invariants given by $L\left( \lambda \right)$.
This shows that the Lax pair \eqref{eq:LaxCET2} does not prove
the na\"ive integrability of the system \eqref{eq:end}.
In this sense the Lax pair \eqref{eq:end} is fake in the sense of
\cite{CalogeroNucci1991,HayButler2013,HayButler2015}.

Now consider the discrete Nahm system \eqref{eq:LPdiscrinv} 
corresponding to the Nahm matrices \eqref{eq:nahmnd}:
\begin{subequations}
    \begin{align}
        T_{1,n} &=
        \begin{pmatrix}
            1&-x_{1,n} &0
             \\ 
             x_{1,n}  &1& x_{4,n}
             \\ 
             0&-x_4 &1
         \end{pmatrix},
        \label{eq:T1ndd}
        \\
        T_{2,n} &=
        \begin{pmatrix}
            1&0& x_{3,n}
            \\ 
            0&1&0
            \\
            -x_{3,n} &0&1
        \end{pmatrix},
        \label{eq:T2ndd}
         \\
         T_{3,n} &=
         \begin{pmatrix}
             1& x_{5,n} &0
             \\ 
             -x_{5,n} & 1 & x_{2,n}
             \\ 
             0&-x_{2,n} &1
         \end{pmatrix}.
        \label{eq:T3ndd}
    \end{align}
    \label{eq:nahmndd}
\end{subequations}
Unfortunately, we obtain the compatibility conditions are overdetermined,
in the sense that we have more compatibility conditions than
independent variables.
For instance, let us consider the first discrete Nahm equation
$T_{1,n+1}-T_{1,n} = h\left( T_{2,n+1}T_{3,n}-T_{3,n+1}T_{2,n} \right)$
writing down the coefficients explicitly:
\begin{subequations}
    \begin{align}
        x_{1,n+1}-x_{1,n}&=h \left( x_{3,n+1}x_{2,n}- x_{5,n}+ x_{5.n+1} \right),
        \\
        x_{3,n+1}-x_{3,n} &=0,
        \label{eq:cet2discrwrongx3p}
        \\
        x_{1,n+1}-x_{1,n} &=h \left( {  x}_{2,{n+1}}{  x}_{3,{n}}-{  x}_{5,{n}}+{  x}_{5,{n+1}} \right), 
        \\
        {  x}_{4,{n+1}}-{  x}_{4,{n}} &=h \left( {  x}_{5,{n+1}}{  x}_{3,{n}} +{  x}_{2,{n}}-{  x}_2,{{n+1}} \right) ,
        \\
        {  x}_{4,{n+1}}-{  x}_{4,{n}} &=h \left( {  x}_{3,{n+1}}{  x}_{5,{n}}+{  x}_{2,{n}}-{  x}_{2,{n+1}} \right).
    \end{align}
    \label{eq:cet2discrwrong}
\end{subequations}
It is clear that \eqref{eq:cet2discrwrongx3p} implies that the discrete
time evolution of $x_{3,n}$ is trivial. So, the discrete Nahm system 
\eqref{eq:cet2discrwrong} cannot be a discretisation of the integrable 
system \eqref{eq:end}.

On the other hand in \cite{PetreraPfadlerSuris2011} it was proven that
the KHN discretisation of the system \eqref{eq:end} exists and it is
algebraically integrable.
More precisely, using the method of the Hirota--Kimura bases \cite{PetreraPfadlerSuris2009}, 
the authors proved that such KHK discretisation preserves all five invariants 
of its continuous counterpart \eqref{eq:end}.
We note that the functionally independent invariants of such KHK discretisation
can be found directly, that is without using the Hirota--Kimura bases,
with the method of \cite{FalquiViallet1993}.
See also \cite{Gubbiotti_Levi70} for an explanation of the method in the 
case of difference equations.

So, to conclude, this example shows that, although very useful
in several case, the Nahm's equations approach is not enough 
to explain integrability of quadratic vector equations in both 
the continuous and the discrete case.

\section*{Acknowledgements}

This research was supported by Dr. M. Radnovi\v{c}'s grant 
DP160101728 and by Prof. N. Joshi and Dr. Milena Radnovi\v{c}'s 
grant DP200100210 from the Australian Research Council.

The author expresses his gratitude to Prof. N. Joshi,
Prof. G. R. W. Quispel and Dr. D. T. Tran
for their helpful discussions during the preparation of this paper.
We thank the anonymous referee, whose comments led to a great
improvement of the paper.

\bibliographystyle{plain}
\bibliography{bibliography}

\end{document}